\documentclass[amssymb,aps,floatfix,floats,preprintnumbers,showpacs,twocolumn]{revtex4-1}
\usepackage{rcs}

\usepackage{graphicx}
\usepackage{subfigure}
\usepackage{dcolumn}
\usepackage{bm}
\usepackage[applemac]{inputenc} 
\usepackage{hyperref} 
\usepackage{sidecap}

\begin{document}

\newcommand{\bsig}{{\bm \sigma}}
\newcommand{\btau}{{\bm \tau}}
\newcommand{\bmu}{{\bm \mu}}
\newcommand{\bs}{{\bm s}}
\newcommand{\bS}{{\bm S}}
\newcommand{\bQ}{{\bf Q}}
\newcommand{\bk}{{\bf k}}
\newcommand{\bq}{{\bf q}}
\newcommand{\bM}{{\bf M}}
\newcommand{\bR}{{\bf R}}
\newcommand{\br}{{\bf r}}
\newcommand{\bt}{{\bf t}}

\newcommand{\bG}{\bar{G}}
\newcommand{\bT}{\bar{T}}
\newcommand{\bSig}{\bar{Sigma}}

\newcommand{\tM}{\tilde{M}}
\newcommand{\hM}{\hat{M}}
\newcommand{\tU}{\tilde{U}}

\newcommand{\bars}{\bar{s}}

\newcommand{\la}{\langle}
\newcommand{\ra}{\rangle}
\newcommand{\dagg}{\dagger}
\newcommand{\ua}{\uparrow}
\newcommand{\da}{\downarrow}

\newcommand{\lc}{\lowercase}

\newcommand{\no}{\nonumber}
\newcommand{\be}{\begin{equation}} 
\newcommand{\ee}{\end{equation}}
\newcommand{\bea}{\begin{eqnarray}} 
\newcommand{\eea}{\end{eqnarray}}

\newcommand{\Q}{($0,0,\frac{2\pi}{c}$)}
\newcommand{\mul}{$\la\mu_{\text{loc}}^2\ra$}
\newcommand{\hch}{\hat{\chi}}

\newcommand{\Bi}{Bi$_2$Se$_3$}

\def\bra#1{{\langle #1 \vert}}
\def\ket#1{{\vert #1 \rangle}}
\def\x#1{{\sigma_{#1}^{x}}}
\def\z#1{{\sigma_{#1}^{z}}}
\def\y#1{{\sigma_{#1}^{y}}}


\title{Inelastic magnetic scattering effect on LDOS  of topological insulators}

\author{Peter Thalmeier and Alireza Akbari}
\affiliation{Max Planck Institute for Chemical Physics of Solids, 01187 Dresden, Germany}
\date{\today}
\begin{abstract}
Magnetic ions such as Fe, Mn and Co with localized spins may be adsorbed on the surface of topological insulators like \Bi. They form scattering centers for the helical surface states which have a  Dirac cone dispersion as long as the local spins are disordered. However, the local density of states (LDOS) may be severely modified by the formation of bound states. Commonly only elastic scattering due to normal and exchange potentials of the adatom is assumed. Magnetization measurements show, however, that considerable magnetic single ion anisotropies exist which lead to a splitting of the local impurity spin states resulting in a singlet ground state. Therefore inelastic scattering processes of helical Dirac electrons become possible as described by a dynamical local self energy of second order in the exchange interaction. The self energy influences bound state formation and leads to significant new anomalies in the LDOS  at low energies and low temperatures which we calculate within T-matrix approach. We propose that they may be used for spectroscopy of local impurity spin states  by  appropriate tuning of chemical potential and magnetic field.
\end{abstract}
\maketitle

\section{Introduction}
\label{sec:introduction}
The warped Dirac cone dispersion and helical spin polarization of surface states in topological insulators (TI) is by now well understood, in particular from ARPES investigations \cite{hsieh:09,chen:10}. These states are topologically protected as long as time reversal invariance is preserved \cite{hasan:10,fu:07}. The latter may be broken by an applied magnetic field  or by adsorption of magnetic adatoms in sufficient concentration such that they exhibit long range order \cite{liu:09}. In this case the Dirac electrons acquire a mass gap proportional to the size of the magnetic field or magnetization. 

Even without long range spin order magnetic adatoms still have a profound effect on the surface states. While there is no gap and at sufficient distance from the impurity the unperturbed Dirac spectrum is recovered, the local density of states (LDOS)  may be severely modified due to formation of low energy bound states at the impurity site \cite{biswas:10}. In addition the  spatial dependence of the DOS away from the impurity may be Fourier transformed to give the quasiparticle interference (QPI) spectrum \cite{guo:10, fu:12, mitchell:12}. It is dominated by the special wave vectors that connect nested Fermi surface pieces not related by backscattering. Therefore the QPI contains considerable information on Fermi surface geometry that may be compared to ARPES results \cite{zhou:09,ye:12}. 

In all cases it is assumed that the scattering mechanism is elastic and no internal degree of freedom may be excited at the impurity site by the scattering. In the case of magnetic impurities it is commonly assumed that the moment is polarized along a fixed direction (the ensuing gap opening due the associated time reversal symmetry breaking is neglected). However this is an oversimplified picture of the real situation. Firstly in zero field and without order the spins are not polarized but fluctuate. Secondly due to the symmetry breaking the directions parallel and perpendicular to the surface plane are not equivalent and therefore single-ion anisotropies for the magnetic moment may be introduced. There is direct experimental evidence for this effect coming from magnetization measurements on Fe adsorbed on \Bi~ \cite{honolka:12} which suggest that the in-plane magnetization is considerably larger than the out-of-plane magnetization. This means that there is an easy-plane spin anisotropy which splits the quantum states of the the $S=2$ Fe impurity spin into three levels. Such anisotropy was also observed in the case of Fe adatoms on graphene \cite{porter:12}. Therefore in second order of the exchange constant there will be {\em inelastic} scattering processes of helical Dirac electrons from those local impurity levels.\\

The internal impurity excitation degrees of freedom can profoundly modify the low energy scattering matrix  and therefore the bound state formation and LDOS. Since the excitation energies of the impurity depend on the field this should also lead to a field dependence of the LDOS even in the case when the  ground state of the magnetic adatom is a singlet. Furthermore the LDOS will be temperature dependent due to the thermal occupation of the impurity levels. One may also expect that these effects depend on the distance of the chemical potential from the Dirac point as long as it is comparable to the splitting energy. We will study these dynamical impurity self energy effects in this work and show that they open interesting perspectives for the STM experiments on topological insulators with magnetic adatoms.\\

We mention that the effect of inelastic scattering on LDOS properties has been investigated before in a completely different context: In superconductors bound states due to  impurity scattering may appear within the superconducting gap \cite{balatsky:06}. The physics of these bound states is strongly modified when the impurities are dynamic, i.e, exhibit a local phonon mode \cite{morr:03} or intramolecular excitations \cite{nyberg:05}. This problem has a certain formal analogy to the one studied here for a completely different and non-superconducting system. Spin-inleastic excitations and their influence on Friedel oscillations for normal metals were investigated in Ref.\onlinecite{fransson:12}.

In Sec.~\ref{sec:costates} we will first discuss the magnetic anisotropy and its connection to quantum states of adatoms with spin. Then we introduce the model for normal and magnetic impurity scattering in Sec.~\ref{sec:impurity}. The scattering matrix and dynamical self energy for the coupled problem are treated in  Sec.~\ref{sec:scattering}. The zero temperature self energy will be calculated analytically in Sec.~\ref{sec:ldosexact} for comparison with numerical results  and the resulting LDOS for various physical regimes will be discussed in Sec.~\ref{sec:discussion}. Finally Sec.~\ref{sec:conclusion} gives the summary and conclusion.

%
\begin{figure}
\raisebox{4.0cm}
{\includegraphics[angle=-90,width=30mm]{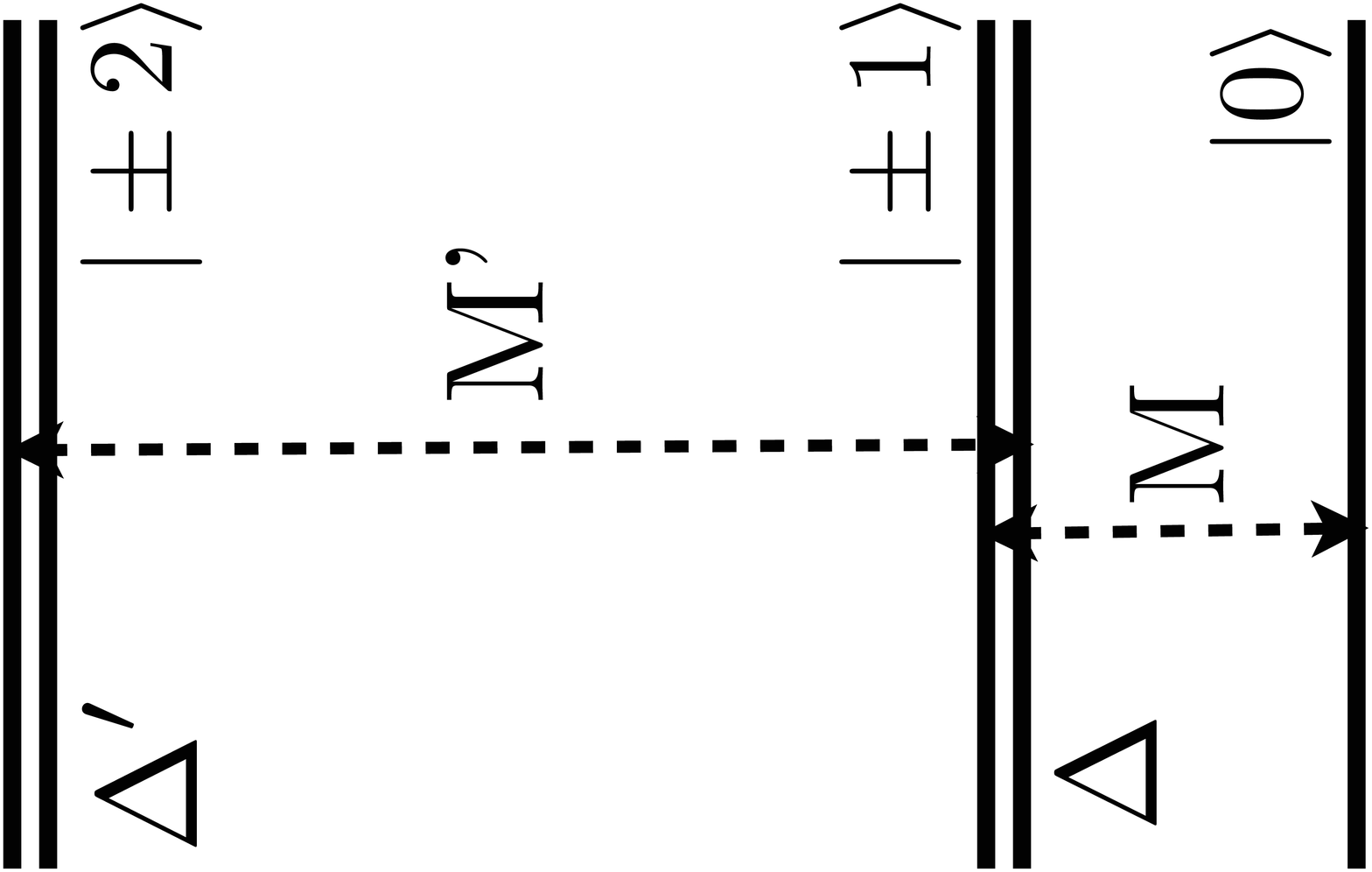}}\hfill
\includegraphics[clip,width=50mm]{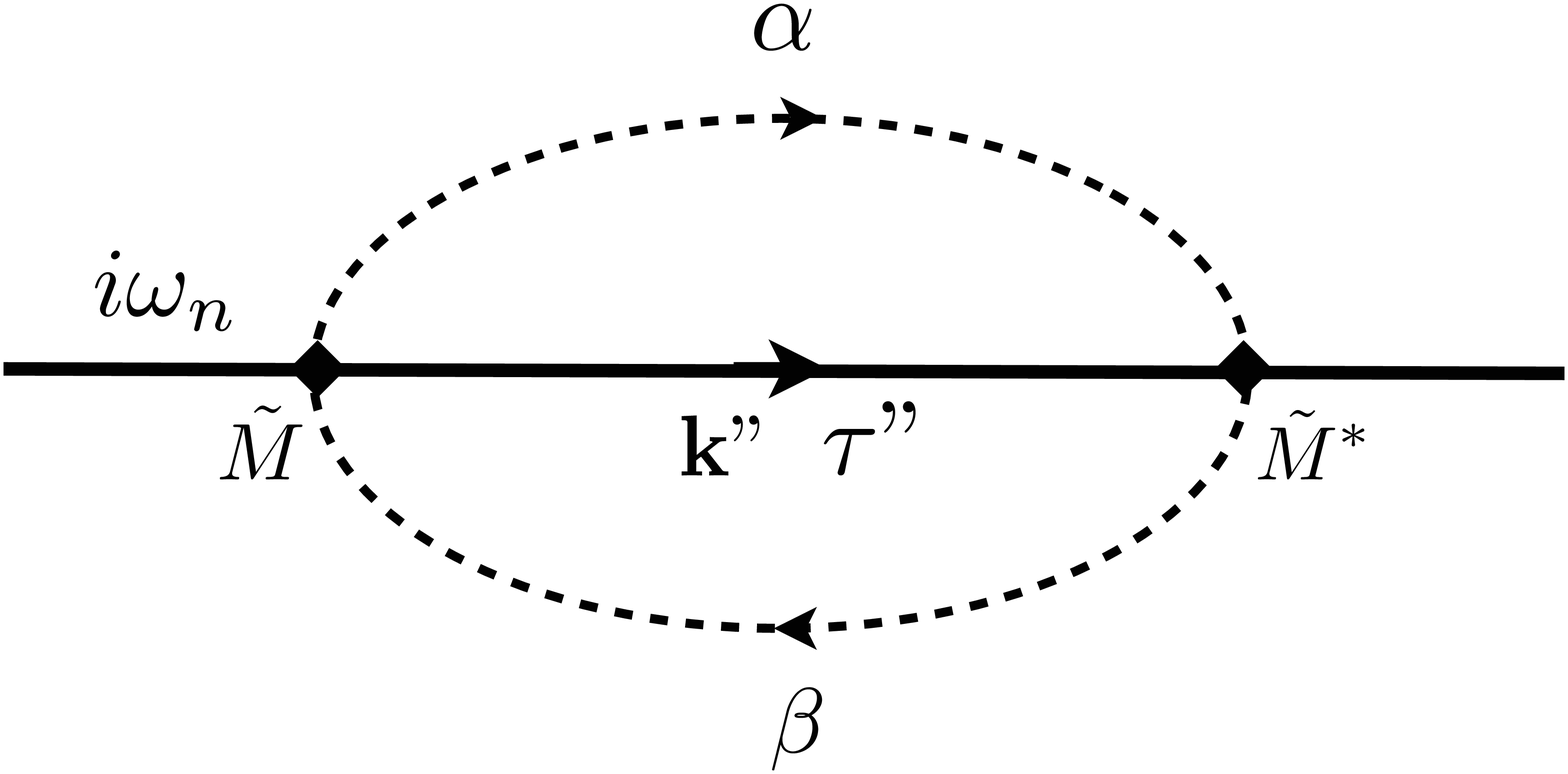}
\caption{Left: Energy level scheme of localized spin states split by the anisotropy term in Eq.~(\ref{eq:anis}) with $\Delta =K_0$ and 
$\Delta'=4K_0$. Right: Second order self energy for Dirac electrons (full line) due to scattering from local spin states $\alpha,\beta$ (broken line).}
\label{fig:Fig1}
\end{figure}
%

\section{Magnetic states of F\lc{e} surface adatoms}
\label{sec:costates}

Various experiments have tested the robustness of TI  surface states against doping with magnetic ions. Bulk doping \Bi~ with 1\% Mn is sufficient to open a magnetic mass gap \cite{chen:10}. Surface deposition of Fe on the (111) selenium surface plane is also reported \cite{wray:11} to lead to massive Dirac dispersion of surface states. On the other hand  these states are insensitive under sufficiently low doping of the same surface  with magnetic Gd atoms \cite{valla:12}. Apparently sufficient surface doping is required to produce a static effective field that leads to the breaking of time reversal symmetry necessary for creating a gap. In addition there is also a more local effect on the electronic structure close to the dopant atom which comes from repeated scattering of Dirac particles at the dopant site and may lead to resonant enhancement of the LDOS at low energies. To describe this effect it is necessary to have a valid model for the possible spin states of the adatom. Due to the spatial symmetry breaking at the surface these states may be split into a sequence of multiplets and lead to anisotropic magnetic properties at the surface. The latter have been reported for Fe doped (111) surfaces of \Bi . From magnetization measurements an in-plane anisotropy was found \cite{honolka:12}. The adsorbed Fe atom is in a $S=2$ high-spin state with moment $m=g_sS\mu_B = 4\mu_B$ and the orbital moment nearly quenched \cite{honolka:12}. The anisotropic single ion spin Hamiltonian is given by
\bea
 {\cal H}_A=(K/S^2)\sum_i S_z(i)^2
 \label{eq:anis}
\eea
Here $K_0=(K/S^2)$ is the magnetic anisotropy constant. It may be directly determined from the full magnetization curves. It also determines the low-field susceptibility anisotropy. The reduced susceptibility $\hat{\chi}=\chi/(g_S\mu_B)^2$ for a field {\bf H} forming an angle $\theta$ with the surface normal $\hat{\bf z}$ is given by
\bea
\hat{\chi}(\theta)&=&\frac{\chi_0}{1+\kappa}
\bigl[\cos^2\theta + (1+\kappa)^2\sin^2(\theta)]\no\\
\kappa&=&\frac{1}{2}(K_0\chi_0)/(g_S\mu_B)^2
\label{eq:sus}
\eea
The dimensionless anisotropy constant $\kappa$ is then determined by $\kappa=\hat{\chi}_\parallel/\hat{\chi}_\perp - 1$ where $\parallel, \perp$ refer to $\theta =\frac{\pi}{2}, 0$ field direction, respectively. In absolute units the anisotropy was determined as
$K_0=0.4$ meV from the magnetization curves \cite{honolka:12}. Due to the positive sign the Fe moment is preferentially aligned with the surface plane. Microcsocpically it splits the five  $|S_z\ra$ local spin ($S=2$)  states into a nonmagnetic singlet ground state and two excited magnetic doublets $|S_z\ra =|0\ra, |\pm1\ra, |\pm 2\ra$ with energies $E_\alpha= 0, \Delta = K_0, \Delta' = 4K_0$  $(\alpha=0,\pm1,\pm2)$ respectively (Fig.~\ref{fig:Fig1}). To treat the scattering of Dirac particles from the impurity spin states we describe the latter by a convenient pseudo fermion representation \cite{abrikosov:65,fulde:72}:
\bea
{\cal H}_A=\sum_{\alpha,i}(E_\alpha+\zeta_i)a_\alpha^\dagg(i)a_\alpha(i)
\label{eq:hanis}
\eea
where the pseudofermion operators $a_\alpha^\dagger(i)$ create the five Fe spin states at impurity site i. Here $\zeta_i$ is a Lagrange parameter used to  project out unphysical states as explained in Appendix \ref{sec:app2}. They correspond to the occupation of two or more different local spin states at the same site.
For Fe on \Bi~ the total splitting is $1.6$ meV which is quite small compared to the surface band cutoff $E_c \simeq 0.25$ eV. In our model calculations we will assume that the splitting may be considerably larger of the order $\Delta/E_c\simeq 0.1$ to exhibit clearly the qualitative effect of inelastic scattering on the LDOS. Such splitting energies seem entirely possible with magnetic 4f adatoms on \Bi.

%
\begin{SCfigure*}
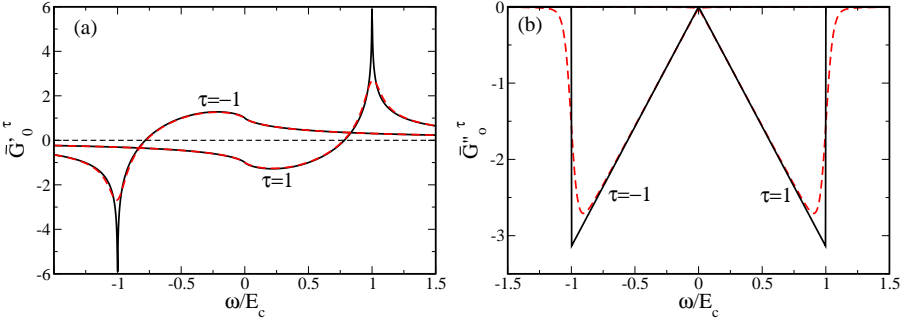

\centering
\includegraphics[width=0.33\textwidth]{Fig2a}
\includegraphics[width=0.33\textwidth]{Fig2b}
\caption{(Color online) (a) real part of conduction electron Green's function $\bG^\tau_0(\omega)'$ from analytical (full line) and numerical (broken line) calculation. The latter is performed with a soft cutoff of width $\gamma/E_c=0.05$ at $E_c$. (b) Corresponding imaginary parts  $\bG^\tau_0(\omega)''$. The DOS of Dirac half cones $(\tau=\pm 1)$ is given by $N_\tau(\omega)/(D_0E_c)=-\frac{1}{\pi}\bG^\tau_0(\omega)''$.
\vspace{0.6cm}}
\label{fig:Fig2}
\end{SCfigure*}
%

\section{Model for normal and magnetic impurity scattering}
\label{sec:impurity}

In this work we will be mostly interested in the effects of inelastic magnetic scattering on the helical surface state LDOS at relatively low energies comparable to the splitting energy $\Delta$ but  much smaller than the bulk gap. Therefore these states may be described by the isotropic Dirac cone Hamiltonian neglecting the warping effect which is negligible in this region. It is given by
\bea
{\cal H}_D=\sum_\bk \Psi_\bk^\dagg\bigl[v(\bk\times\bsig)\cdot\hat{\bf z}\bigr]\Psi_\bk
=\sum_\bk\Psi_\bk^\dagg h_D^\bk\Psi_\bk
\label{eq:hdirac}
\eea
with $\Psi^\dagg_\bk=(c_{\bk\ua}^\dagg,c_{\bk\da}^\dagg)$ denoting the surface states. It is diagonalized by the helical eigenstates 
$\Phi_\bk^\dagg=(\gamma_{\bk +}^\dagg,\gamma_{\bk -})$ where $\tau=\pm 1$ is the helicity. Explicitly $h_D^\bk=E_\bk\tau_z$ with the helical eigenvalues $E_{\bk \tau}=\tau E_\bk$ given by $E_\bk=v|\bk|$ where $v$ is the velocity of Dirac particles. Here $\tau=\pm1$ denote the upper and lower Dirac cone respectively. The eigenstates are obtained by 
\bea
\Phi_\bk=P_\bk^\dagg\Psi_\bk; \;\;\;
P_\bk=\left(\matrix
{1& ie^{-i\phi_\bk} \cr
 -ie^{-i\phi_\bk} &  1}\right)
 \label{eq:helical}
\eea
where $\phi_\bk=\tan^{-1}(k_y/k_x)$ is the azimuthal angle of {\bf k}. 

The scattering of these surface states from normal and magnetic impurities is described by contact interactions of strength $U_0$ and $J_0$ at the impurity sites $\bR_i$:
\bea
U(\br-\bR_i)&=&U_0\delta(\br-\bR_i)\no\\
V(\br-\bR_i)&=&\frac{1}{2}J_0\bsig\cdot\bS\delta(\br-\bR_i)
\label{eq:impurity}
\eea
Their Fourier transforms are momentum independent with $U_{\bk\bk'}=U_0$ and $V_{\bk\bk'}=\frac{1}{2}J_0\bsig\cdot\bS$. The total scattering Hamiltonian $H_I=H_U+H_J$ is then given by
\bea
{\cal H}_U&=&U_0\sum_i\sum_{\bk\bq\sigma}c_{\bk+\bq\sigma}^\dagg c_{\bk\sigma'}e^{i\bq\bR_i}\\
{\cal H}_J&=&\frac{1}{2}J_0\sum_{i\bk\bq}\sum_{\alpha\beta\sigma\sigma'}M_{\alpha\beta}^{\sigma\sigma'}
a_\alpha^\dagg(i)a_\beta(i)c_{\bk+\bq\sigma}^\dagg c_{\bk\sigma'}e^{i\bq\bR_i}\no
\label{eq:himpurityc}
\eea
The matrix elements result from the pseudo-fermion representation for spin operators which may be expressed as
\bea
\bS_i&=&\sum_{\alpha\beta}\la \alpha|\bS_i|\beta\ra a_\alpha^\dagg(i)a_\beta(i)\no\\
M_{\alpha\beta}^{\sigma\sigma'}&=&\la \alpha|\bS_i|\beta\ra\cdot\bsig_{\sigma\sigma'}
\label{eq:pseudo}
\eea
Here $\alpha,\beta$ denote the split impurity spin states. The total model Hamiltonian is then given by
${\cal H} ={\cal H}_A + {\cal H}_D + {\cal H}_I$. It is necessary to transform {\cal H} to the helical basis states created by 
$\gamma_{\bk\tau}^\dagg$ using the unitary transformation in Eq.~(\ref{eq:helical}). We separate the mean field term 
$\sim \la {\bf S}\ra$ of the exchange part and add it to the potential scattering
Then we obtain
\bea
{\cal H}_{I}^{MF}&=&\sum_i\sum_{\bk\bq\tau}\tU^{\tau\tau'}(\bk,\bk+\bq)
\gamma_{\bk+\bq\tau}^\dagg \gamma_{\bk\tau'}e^{i\bq\bR_i}\\
{\cal H}_J&=&\frac{1}{2}J_0\sum_i\sum_{\bk\bq\tau\tau'}\tM_{\alpha\beta}^{\tau\tau'}a_\alpha^\dagg(i)a_\beta(i)\no
\gamma_{\bk+\bq\tau}^\dagg\gamma_{\bk\tau'}e^{i\bq\bR_i}
\label{eq:himpurityg}
\eea
For the mean field part we have, denoting $\bk'=\bk+\bq$
\bea
\tU^{\tau\tau'}(\bk\bk')=U_0\tU_0^{\tau\tau'}(\bk\bk')+U_1\tU_1^{\tau\tau'}(\bk\bk')
\label{eq:umf}
\eea
The explicit form of matrix elements $\tU^{\tau\tau'}(\bk\bk')$ and $\tM^{\tau\tau'}_{\alpha\beta}(\bk\bk')$  is given in Appendix \ref{sec:app1}.
Assuming that the mean field spin polarization is oriented along z  the corresponding scattering strength is given by $U_1=(J_0/2)\la S_z\ra$. 
The spin polarization can appear as result of spontaneous FM order of impurity spins for sufficient surface coverage or through an applied field $H$. We note that in the present model this is $\sim H^2$ because the impurity spin ground state is a nonmagnetic singlet.

\section{Self energy and scattering matrix}
\label{sec:scattering}

The mean field part of the exchange scattering contained in ${\cal H}_I^{MF}$ is of first order in $J_0$ and therefore a static contribution. However the second order dynamic self energy due to  ${\cal H}_J$ will exhibit frequency dependence determined by the level splittings of the localized spin (Fig~\ref{fig:Fig1}). The evaluation of the corresponding self energy diagram in this figure is described in Appendix \ref{sec:app2}.
The expression for the self energy given in Eq.~(\ref{eq:sig2a}) for general level scheme will  be simplified to the case which is physically relevant here.  Firstly we note that the dipolar matrix elements lead only to two inelastic transitions and only one from the ground state. Explicitly $\la \pm1| S_\pm |0\ra \equiv M= \sqrt{6}$ and 
$\la \pm 2| S_\pm |\pm1\ra \equiv M'= 2$. Secondly for $T < \Delta$ the $M'$ transition may be neglected since it starts from an excited state. Including only the M transition the self energy reduces to
\bea
&&\Sigma_2(\bk,\tau,\bk'\tau';i\omega_n)=(J_0/2)^2p(T)\sum_{\bk''\tau''}
\hM_{\bk\bk'}^{\tau\tau'}(\bk''\tau'')\no\\
&&\times\Bigl[\frac{1-f(\epsilon_{\bk''\tau''})+n_B(\Delta)}
{i\omega_n-\epsilon_{\bk''\tau''}-\Delta}
+\frac{f(\epsilon_{\bk''\tau''})+n_B(\Delta)}
{i\omega_n-\epsilon_{\bk''\tau''}+\Delta}\Bigl]
\label{eq:sig2c}
\eea
where $p(T)=2tanh(\Delta/2T)/[3-tanh(\Delta/2T)]$ is the occupation difference of singlet ground state and first excited doublet and $f(\omega)$ and $n_B(\omega)$ are the Fermi and Bose functions respectively. The matrix elements $\hM_{\bk\bk'}^{\tau\tau'}(\bk''\tau'')$  and Matsubara frequencies $\omega_n$ are defined in Appendix \ref{sec:app2}.
Using Eq.~(\ref{eq:umf}) the total scattering potential for Dirac electrons entering into the T- matrix calculation is then given by
%
\begin{figure}
\includegraphics[width=70mm]{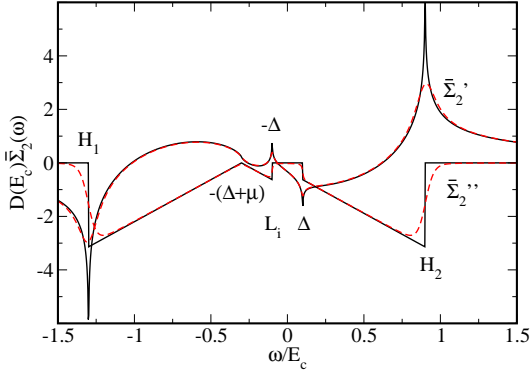}
\caption{(Color online) Real and imaginary part  of self energy $\bar{\Sigma}_2(\omega)$ at $T=0$ for $\mu/E_c=0.2$; $\Delta/E_c=0.1$ and $(\lambda_0, \lambda) =(0,1)$ . The singular points given in Eq.~(\ref{eq:singular}) at high energies close to the surface band edge ($H_1,H_2$)
and low energies close to the Dirac point $(L_i, i=1-3)$ are indicated. Full line corresponds to excact result using Eqs.~(\ref{eq:sig1},\ref{eq:sig2}) with sharp cutoff at $E_c$ and dashed line to numerical integration of Eq.~(\ref{eq:sig2av}) with soft cutoff $\gamma/E_c=0.05$ and imaginary part
$\eta =0.5\cdot10^{-3}$.}
\label{fig:Fig3}
\end{figure}
%
%
\bea
\tU_t^{\tau\tau'}(\bk\bk';i\omega_n)=\tU^{\tau\tau'}(\bk\bk')+\Sigma_2(\bk,\tau,\bk'\tau';i\omega_n)
\label{eq:utot}
\eea
It is now depending on the energy of conduction electrons due to the second order exchange self energy.\\

The local density of states of conduction electrons at the impurity site can be obtained from the expression \cite{liu:12}
\bea
\delta N_\tau(\omega)=-\frac{1}{\pi}\sum_{\bk\bk'}tr\bigl[G_0(\bk,\omega)T(\bk\bk',\omega)G_0(\bk',\omega)\bigr]
\label{eq:ldos}
\eea
where the T-matrix due to the scattering potential $\tU_t$ satisfies the equation
\bea
T(\bk,\bk',\omega)&=&\tU_t(\bk,\bk';\omega)+\\
&&\sum_{\bk''}\tU_t(\bk,\bk'';\omega)G_0(\bk'',\omega)T(\bk'',\bk',\omega)\no
\label{eq:tmat}
\eea
The solution of this equation is simplified by the following observation. If we neglect the warping term in the Dirac Hamiltonian
and denote $ \bk$ by its polar coordinates $(k,\phi)$ then in Eqs.~(\ref{eq:tmat},\ref{eq:ldos}) $G_0(k,\omega)$ and $G_0(k',\omega)$ depend only on the moduli and $\tU_t(\phi,\phi',\omega)$ only on the angles of the wave vectors. Then the summation in these equations may be done separately for the Green's functions and the T-matrix. Replacing $\sum_\bk \rightarrow (A/2\pi^2)\int dkd\phi$ we obtain the averages
\bea
\bG_0(\omega)&=&\frac{A}{2\pi}\int dk G_0(k,\omega)\no\\
\bar{U}_t(\omega)&=&\frac{1}{2\pi^2}\int d\phi d\phi' \tU_t(\phi,\phi',\omega)\\
\bT(\omega)&=&\frac{1}{2\pi^2}\int d\phi d\phi' T(\phi,\phi',\omega)\no
\label{eq:average}
\eea
From Eqs.(\ref{eq:amat},\ref{eq:tm}) we can see that in $\bar{U}_t$ all $\phi, \phi'$ dependent terms average to zero. Therefore
\bea
\bar{U}_t(\omega)=[U_0+\bar{\Sigma}_2(\omega)]\sigma_0+U_1\sigma_z
\label{eq:uav1}
\eea
where $\sigma_0$ is the unit and $\sigma_z$ a Pauli matrix. Then Eq.(\ref{eq:tmat}) may be written as
\bea
\bT(\omega)&=&\bar{U}_t(\omega)[1-\bar{U}_t(\omega)\bG_0(\omega)]^{-1}
\label{eq:tav}
\eea
%
%
\begin{figure}
\includegraphics[width=70mm]{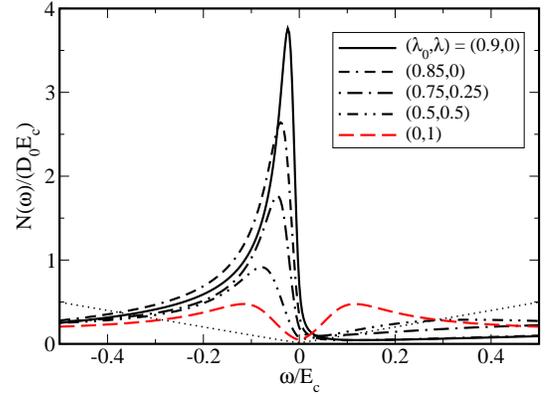}
\caption{(Color online) LDOS for $\mu=0$,  $\Delta =0$ and various scattering strengths $(\lambda_0,\lambda)$. Unperturbed $D(\omega)=D_0|\omega|$ is shown as dotted line. Numerical evaluation with  $\eta/E_c=0.5\cdot 10^{-3}$ is used.}
\label{fig:Fig4}
\end{figure}
%
%
Using helical eigenstates ($\tau=\pm 1$) we may evaluate the trace for the LDOS $\delta N(\omega)=\sum_\tau\delta N_\tau(\omega)$. For each helicity channel it is finally given by
\bea
\delta N_\tau(\omega)=-\frac{1}{\pi}\frac{\bar{U}^\tau_t(\omega)\bG^\tau_0(\omega)^2}
{1-\bar{U}^\tau_t(\omega)\bG^\tau_0(\omega)}
\label{eq:ldosav}
\eea
where 
\bea
\bar{U}^\tau_t(\omega)=U_0+\tau U_1+\bar{\Sigma}_2(\omega)
\label{eq:uav2}
\eea
is the total energy dependent scattering potential at the impurity site and $(i\omega_n\rightarrow \omega+i\eta)$
\bea
\label{eq:sig2av}
\bar{\Sigma}_2(i\omega_n)&=&\frac{1}{2}M^2J_0^2p(T)\sum_{\tau''}\int_0^{E_c}D(E)\\
&&\times\Bigl[\frac{1-f(\epsilon_{\tau''})+n_B(\Delta)}
{i\omega_n-\epsilon_{\tau''}-\Delta}
+\frac{f(\epsilon_{\tau''})+n_B(\Delta)}
{i\omega_n-\epsilon_{\tau''}+\Delta}\Bigl]\no
\eea

with $\epsilon_\tau=\tau E-\mu$.
Here, due to vanishing warping a change to integration over the DOS has been used. The latter is
given by $D(E) = D_0E$ with $D_0=A/(2\pi v^2)$ where
 $A = a^2$  is the area of the surface unit cell related to the
Wigner-Seitz radius $k_0$ of the surface BZ by $k_0^2=4\pi/A$ or $k_0=(2\pi/a)/\sqrt{\pi}$. It is associated with
the zone-boundary extrapolated cone energy according to $E_0=vk_0$.
%
\begin{figure}
\includegraphics[width=70mm]{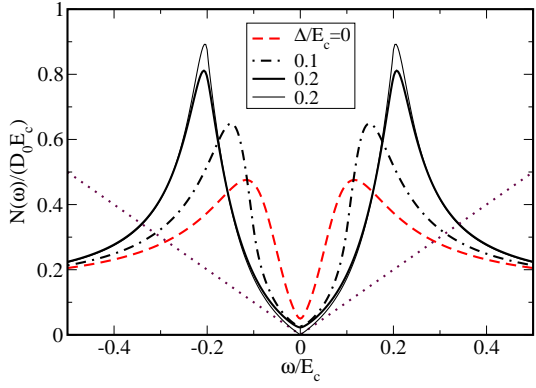}
\caption{(Color online) LDOS at $T=0$ for $\mu=0$ and scattering strengths $(\lambda_0,\lambda)=(0,1)$ for  various $\Delta$. For $\Delta/E_c=0.2$ the  thick line is from numerical calculations $(\eta/E_c=0.5\cdot 10^{-3})$ and the thin line from exact analytical results. Unperturbed $D(\omega)=D_0|\omega|$ is shown as dotted line.}
\label{fig:Fig5}
\end{figure}
%
%
\begin{figure}
\vspace{0.5cm}
\includegraphics[width=70mm,clip]{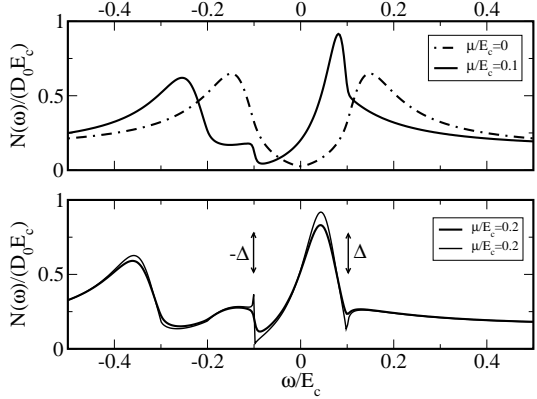}
\caption{LDOS at $T=0$ for $\Delta/E_c=0.1$, scattering strengths $(\lambda_0,\lambda)=(0,1)$ and various $\mu$.
In the lower figure thick line represent numerical results 
 $(\eta/E_c=0.5\cdot 10^{-3})$  and thin line the exact analytical results. Arrows indicate position of $L_{2,3}$ singularities in
 self energy due to local spin excitations.}
\label{fig:Fig6}
\end{figure}
%
The cutoff wave vector is $k_c=E_c/v\ll k_0$ and
the cutoff energy $E_c\ll E_0$ is of the order of half of the bulk band gap ($\approx 0.25$ eV for \Bi) where helical surface states
merge into the bulk states. For numerical calculations we introduce a soft cutoff function $f_c(\omega)=\frac{1}{2}[1-\tanh((E-E_c)/\gamma)]$ with $\gamma$ denoting the width of the cutoff around $E_c$. This ensures a finite width to the bound states around
the cutoff energy and also improves the convergence of numerical integration.
Potential scattering and second order self energy may be characterized by dimensionless
coupling constants $\lambda_0$ and $\lambda$ that are, respectively, given by
\bea
\lambda_0&=&D(E_c)U_0\no\\
\lambda(T)&=&\frac{1}{2}M^2[J_0D(E_c)]^2p(T)
\eea
Here $\lambda$ depends on temperature due to the occupation factor from impurity spin states. Our interest will be focused on the case when bound state effects in the LDOS appear and these interaction parameters will be chosen accordingly in the discussion.
The first order exchange scattering potential $U_1=\frac{1}{2}J_0\la S_z\ra$ will not be considered because it vanishes due to the impurity singlet ground state,
i.e. $\la 0|S_z|0\ra=0$ and is only weakly induced with $ \la S_z\ra\approx H^2$ by a magnetic field.

The above expressions in Eqs.~(\ref{eq:ldosav},\ref{eq:uav2},\ref{eq:sig2av}) can now be used to caculate the LDOS numerically as function of frequency and temperature. A finite imaginary part $\eta$ for the frequency will be used for the integration.

\section{Analytical calculation  at T=0}
\label{sec:ldosexact}

Instead of presenting immediately  numerical results as is frequently done within this context, we first develop the theory much further to a closed analytical solution for T=0 which is possible here due to the restriction to isotropic Dirac states with linear quasiparticle DOS.
For zero temperature we have $n_B(\Delta)=0$, $f(\epsilon_\tau)=1-\Theta(\mu-\epsilon_\tau)$ and p(0)=1. We then may continue to the real axis and evaluate the integrals for $\bG^\tau_0(\omega)=\bG^\tau_0(\omega)'+i\bG^\tau_0(\omega)''$ and $\bar{\Sigma}_2(\omega)=\bar{\Sigma}_2(\omega)'+i\bar{\Sigma}_2(\omega)''$. The former is given by
\bea
\bG^\tau_0(\omega)'&=&D_0\bigl[-\tau E_c-(\omega+\mu)\ln|1-\tau\frac{E_c}{\omega+\mu}|\bigr]\\
\bG^\tau_0(\omega)''&=&-\pi D_0\tau(\omega+\mu)\Theta[\tau(\omega+\mu)]\Theta[E_c-\tau(\omega+\mu)]\no
\label{eq:gfunc}
\eea
Here $\Theta(x)$ is the Heaviside function. Likewise we may obtain the real and imaginary part of the self energy. Defining
$\bar{\Sigma}_2(\omega)=\frac{1}{2}M^2J_0^2p(0)\hat{\Sigma}_2(\omega)=(\lambda/D(E_c)^2)\hat{\Sigma}_2(\omega)$ we get:
\begin{widetext}
\bea
\hat{\Sigma}_2(\omega,\mu)'=D_0\Bigl\{
&&\bigl[(\omega-\Delta+\mu)\bigl(\ln|1+\frac{\mu}{\omega-\Delta}|-\ln|1-\frac{E_c}{\omega-\Delta+\mu}|\bigr)
-(\omega+\Delta+\mu)\ln|1+\frac{E_c+\mu}{\omega+\mu}|\bigr]\Theta(-\mu)\no\\
+&&\bigl[(\omega+\Delta+\mu)\bigl(\ln|1+\frac{\mu}{\omega+\Delta}|-\ln|1+\frac{E_c}{\omega+\Delta+\mu}|\bigr)
-(\omega-\Delta+\mu)\ln|1-\frac{E_c-\mu}{\omega-\mu}|\bigr]\Theta(\mu)
\Bigr\}
\label{eq:sig1}
\eea
\bea
\hat{\Sigma}_2(\omega,\mu)''&=&-\pi D_0\Bigl\{
(\omega-\Delta+\mu)\Theta(\omega-\Delta)
\bigl[\Theta(\omega-\Delta+\mu)\Theta(E_c-\omega+\Delta-\mu)-\Theta(-\omega+\Delta-\mu)\Theta(E_c+\omega-\Delta+\mu)\bigr]\no\\
&&+(\omega+\Delta+\mu)\Theta(-\omega-\Delta)
\bigl[\Theta(\omega+\Delta+\mu)\Theta(E_c-\omega-\Delta-\mu)-\Theta(-\omega-\Delta-\mu)\Theta(E_c+\omega+\Delta+\mu)\bigr]\Bigr\}
\label{eq:sig2}
\eea
\end{widetext}

For chemical potential lying at the Dirac point ($\mu=0$) these expressions simplify to

\bea
\hat{\Sigma}_2(\omega,\mu)'=-D_0\bigl[
&&(\omega-\Delta)\ln|1-\frac{E_c}{\omega-\Delta}|\no\\
&&+(\omega+\Delta)\ln|1+\frac{E_c}{\omega+\Delta}|\Bigr]
\label{eq:sig10}
\eea
and
\bea
\hat{\Sigma}_2(\omega,\mu)''&=&-\pi D_0\bigl[
(\omega-\Delta)\Theta(\omega-\Delta)  \Theta(E_c-\omega+\Delta)\no\\
&&-(\omega+\Delta)\Theta(-\omega-\Delta)\Theta(E_c+\omega+\Delta)\bigr]
\label{eq:sig20}
\eea
In this case the self energy fulfils the symmetry relations
$\hat{\Sigma}_2(-\omega,\mu)'=-\hat{\Sigma}_2(\omega,\mu)'$ and
$\hat{\Sigma}_2(-\omega,\mu)''=\hat{\Sigma}_2(\omega,\mu)''$.\\

The second order self energy has singularities at high energies close to
the surface band edges $\omega\approx\pm E_c$ denoted by $H_{1,2}$
and at low energies close to the Dirac point denoted by $L_{1-3}$. The positions of
these singularities are
\bea
\omega(L_1)&=&(\Delta-\mu)\Theta(-\mu)-(\Delta+\mu)\Theta(\mu)\no\\
\omega(L_2)&=&\Delta\no\\
\omega(L_3)&=&-\Delta\no\\[0.2cm]
\omega(H_1)&=&-E_c-(\Delta+\mu)\no\\
\omega(H_2)&=& E_c+(\Delta-\mu)
\label{eq:singular}
\eea

Because they occur in the denominator of the T-matrix they have a direct influence
on the positions of its bound state poles and therefore on the peak positions and
anomalies of the LDOS. Only the $\omega(L_1)$ position depends explicitly on the chemical
potential, it is positive for $\mu <0$ and negative for $\mu > 0$ $(\Delta > 0)$. This singularity
is the weakest because it appears only in the slope of the self energy.

\section{Discussion of LDOS results}
\label{sec:discussion}

In this section  we discuss the results of the theory outlined above. We focus on the effect of second order inelastic exchange scattering on
the LDOS as given by Eq.~(\ref{eq:ldosav}). In this expression both real and imaginary part of the Green's function and self energy appear.
In the whole discussion we will ignore the first order exchange scattering $U_1$ because it is $\sim \langle S_z\rangle$ which vanishes for
the singlet ground state (Fig.~\ref{fig:Fig1}), even in finite field it will increase only $\sim H^2$.

The Green's functions are presented in Fig.~\ref{fig:Fig2}. In Fig.~\ref{fig:Fig2}(a) the real part for electrons in chiral states $\tau=\pm 1$ is shown. Full and dashed
lines correspond to analytical and numerical results respectively. The agreement is close keeping in mind that exact results are for a sharp cutoff and $\omega$ real whereas in the numerical calculation a soft cutoff with $\gamma/E_c$=0.05 around the band edge and a finite imaginary part $\eta$ of the frequency  are employed. Fig.~\ref{fig:Fig2}(b) shows the imaginary part which is
equal to $-\pi N_\tau(\omega)/(D_0E_c)$ where $N_\tau$ is the DOS of the individual Dirac cones.

The T=0 self energy for the general case with both $\mu, \Delta$ non-zero and different is shown in Fig.~\ref{fig:Fig3}. Again there is an excellent agreement between analytical (full lines) and numerical results, even on the details around the singular points $L_{1-3}$, $H_{1,2}$ which may be clearly identified. This leads to a confidence that the features in the LDOS that originate in the self energy due to inelastic scattering will be reliably obtained.
We should note that the  self energy singularities $H_{1,2}$ close to the band edges and also those at $\pm E_c$ in the Green's functions (Fig.~\ref{fig:Fig2}) are to some degree artificial, because they are generated by the sharp or soft cutoff at the surface band edges. This will also lead to artificial bound states at or outside the band edges and corresponding peaks in the model LDOS as mentioned in Ref.~\cite{biswas:10}. In reality for $\omega\approx \pm E_c$ the surface bands merge into the bulk states which will also show a scattering from the impurity potentials. Therefore the transition around $E_c$ will be very gradual meaning that the high energy singularities will be much weaker. Their proper analysis would have to include the surface states as well as the bulk states on an equal footing such as described  in Refs.~\cite{hao:11}. This is beyond the present effective surface band model. Therefore in the following when discussing the LDOS we will focus on the low energy part with $\omega/E_c\leq 0.5$ which are not much influenced by the details of the cutoff procedure and the merging of surface and bulk states.

First we consider the LDOS according to Eq.(\ref{eq:ldosav}) for the symmetric case $\mu=\Delta=0$ but with variable relative strengths of potential ($\lambda_0$) and exchange ($\lambda$) scattering. The results are shown in Fig.~\ref{fig:Fig4} for various $(\lambda_0,\lambda)$. For pure potential scattering $\lambda_0\approx 1 > 0$  (full line) a strong bound state peak develops slightly below the Dirac point because the real part $1-U_0G_0^-(\omega)'$  vanishes while the imaginary part  in Eq.(\ref{eq:ldosav}) is still small. For the opposite sign of the scattering potential $\lambda_0\approx -1$ the bound state pole develops for the $\tau=+$ helicity and lies slighly above the Dirac point.
If we turn on the exchange scattering the real part  $1-\tU_t(\omega)G_0^-(\omega)'$ vanishes further away from the Dirac point because $\bar{\Sigma_2}(0)'=0$  in the symmetric case. Then the absolute value of the imaginary part will be larger. Therefore the bound state signature in the LDOS is shifted to lower energy, broadened and suppressed as seen in Fig.~\ref{fig:Fig4}. Finally for pure exchange scattering $\lambda =1$ ($\lambda_0=0$) the broad bound state peak appears on both sides of the Dirac cone because  $\bar{\Sigma}_2(\omega)'$  changes sign at $\omega=0$. In the following we will mostly discuss this case because for 3d/4f adatoms the exchange scattering should dominate due to the effect of strong intra-shell correlations.
%
\begin{figure}
\centering
\subfigure[]
{\includegraphics[width=0.46\linewidth]{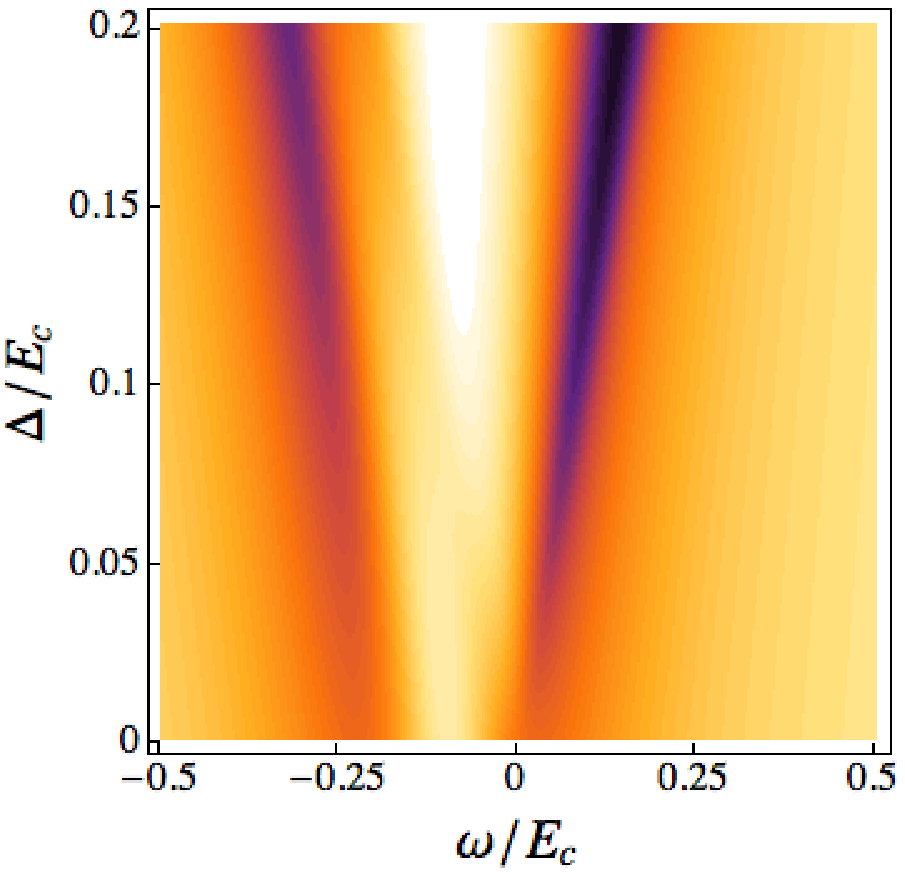}}
\subfigure[]
{\includegraphics[width=0.46\linewidth]{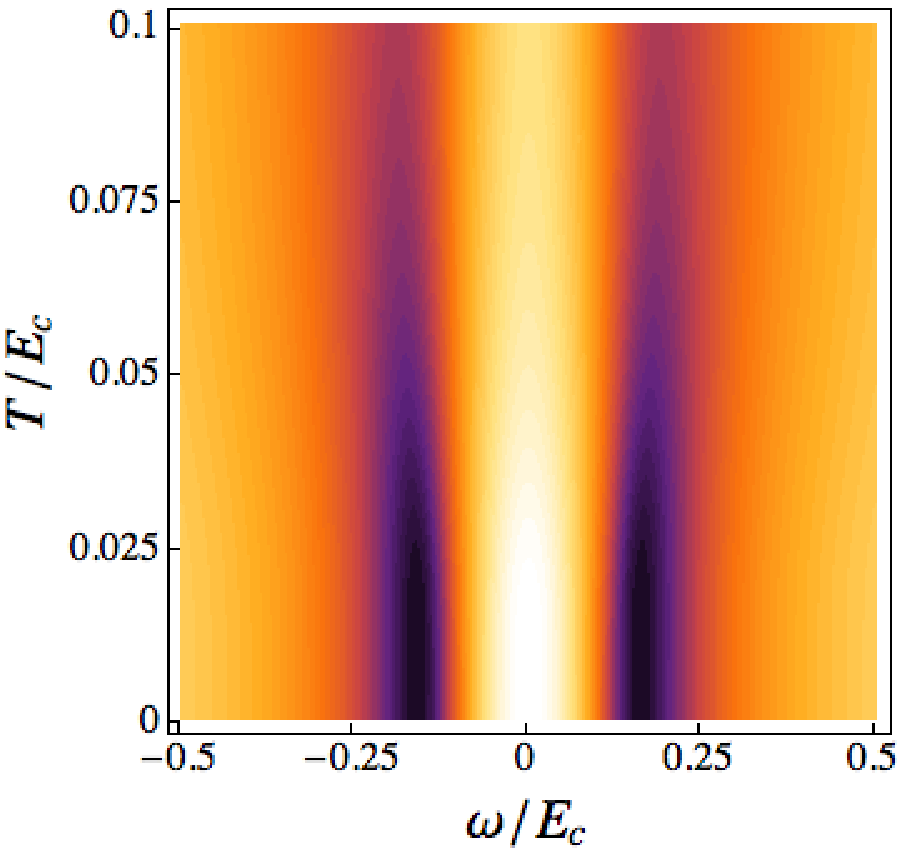}}
\caption{(Color online) Contour plot of LDOS with $\lambda_0=0, \lambda(T=0)=1$. (a)  In $\omega-\Delta$ plane with chemical potential at $\mu/E_c=0.1$ . Increasing $\Delta$ leads to  separation and increasing prominence of bound state peaks with asymmetric intensity for non-zero $\mu$. (b) In $\omega-T$ plane for symmetric case $\mu=0$ and $\Delta/E_c =0.1$.}
\label{fig:Fig7}
\end{figure}
%
The effect of the inelastic local spin excitations on the LDOS for $\mu=0$ is shown in Fig.~\ref{fig:Fig5}. The Dirac point in the self energy now widens to a finite interval $[-\Delta,\Delta]$ according to Eq.~(\ref{eq:sig20}) (see also in the general case of  Fig.~\ref{fig:Fig3}). Therefore the bound states are pushed symmetrically further away from the Dirac point with increasing $\Delta/E_c$ and growing sharper. Remarkably however, the self energy singularities $L_{2,3}$ do no leave a signature in the LDOS at $\pm\Delta$. This happens because for $\mu=0$ they are also of the weak type (only in slope) as $L_1$ is for arbitrary $\mu$. 

The direct signature of inelastic spin excitations finally appears in the LDOS if we consider the most general case with $\mu,\Delta$ different from zero presented in Fig.~\ref{fig:Fig6}. Then $L_{1,2}$ turn into strong logarithmic singularities  (Fig.~\ref{fig:Fig3}) that can also be seen as anomalies in the LDOS . They are stationary in $\omega$ as function of $\mu$ and show increasing prominence for larger $\mu$. This may be understood from the terms like $\sim (\omega+\Delta+\mu)\bigl(\ln|1+\frac{\mu}{\omega+\Delta}|\bigr)$ in Eq.~(\ref{eq:sig1}). For $\omega\simeq-\Delta$ they reduce to $\mu\ln|\frac{\mu}{\omega+\Delta}|$ which is a logarithmic singularity at $-\Delta$ whose weight is only finite for non-zero $\mu$. Indeed while the effect of this singularity in the LDOS is absent in  Fig.~\ref{fig:Fig5} where $\mu=0$ its influence grows with non-zero $\mu$ in Fig.~\ref{fig:Fig6}.
In fact for $\mu/E_c=0.2$ a third broad peak in the spectrum appears right below $-\Delta$ which is connected with this inelastic transition. The one at $\Delta$ is less prominent but still visible. for $\mu<0$ the situation would simply be reversed. Since the chemical potential may be tuned by applying a gate voltage \cite{chen:10b,checkelsky:11} the evolution of these inelastic scattering features in the surface LDOS should be observable. The main resonance peaks (asymmetric for $\mu\neq 0$) are already present for $\Delta=0$ but their distance increases with growing $\Delta$ where the asymmetry is also more pronounced as shown in the LDOS contour plot in Fig.~\ref{fig:Fig7}a.
%
\begin{figure}
\includegraphics[width=70mm]{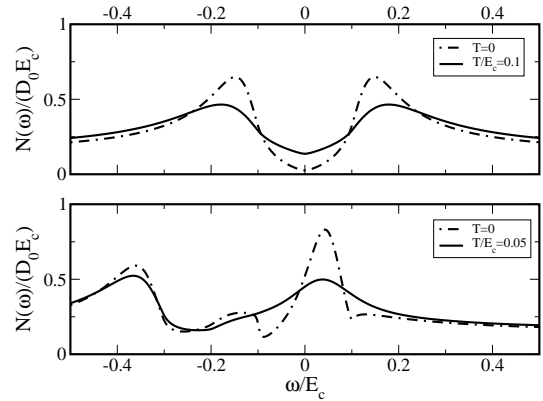}
\caption{Temperature dependence of LDOS (numerical calculations for $\Delta/E_c=0.1$ with $\eta/E_c=0.5\cdot 10^{-3})$. 
Top: $\mu/E_c=0$. Bottom: $\mu/E_c=0.2$. }
\label{fig:Fig8}
\end{figure}
%

We now discuss the temperature dependence of the effects caused by the inelastic local spin excitations on the LDOS. There are two temperature effects: Firstly the effective exchange coupling $\lambda(T)$ for $T>0$ is proportional to the occupation factor $p(T)$ which decreases monotonically from one at
$T\ll\Delta$ to zero for $T\gg\Delta$. Therefore it reduces the size of the self energy due to inelastic scattering for larger temperatures. If we assume that potential scattering is also present then a change of their relative weight as illustrated in Fig.~\ref{fig:Fig4} can be achieved by varying temperature. Secondly the self energy will depend on temperature through the Fermi functions which tend to smooth the singularities for increasing temperatures. Both effects will suppress the bound state features in the LDOS spectrum. The global behaviour is shown in the LDOS contour plot of  Fig.~\ref{fig:Fig7}b for the symmetric case $\mu =0$. While the position of the resonance peaks remain almost constant their intensity is rapidly diminished with increasing temperature. Cuts at constant temperature are shown in Fig.~\ref{fig:Fig8} for $\Delta/E_c=0.1$. The top panel corresponds to $\mu=0$ and $T/E_c=0.1$  and the lower panel to $\mu/E_c=0.2$ and $T/E_c=0.05$ (full lines). For  comparison the $T=0$ result is also shown (dashed lines). Apparently in the case  $\mu > 0$ a moderate temperature is sufficient to smooth out the additional bound state structure resulting from the $L_3$- singularity at $-\Delta$ (cf. Fig.~\ref{fig:Fig6}).

We also address the influence of an external magnetic field on the LDOS spectrum  for $T=0$ which should be considerable a low energies $|\omega| \simeq \Delta$. In the following we consider the easy-plane case ($K_0 >0$) as before and in addition the easy-axis case ($K_0 <0$) for the local moment anisotropy in Eq.~(\ref{eq:anis}) because their LDOS shows interesting differences. They correspond to magnetization or susceptibility enhanced  parallel or perpendicular to the surface respectively.

\begin{figure}
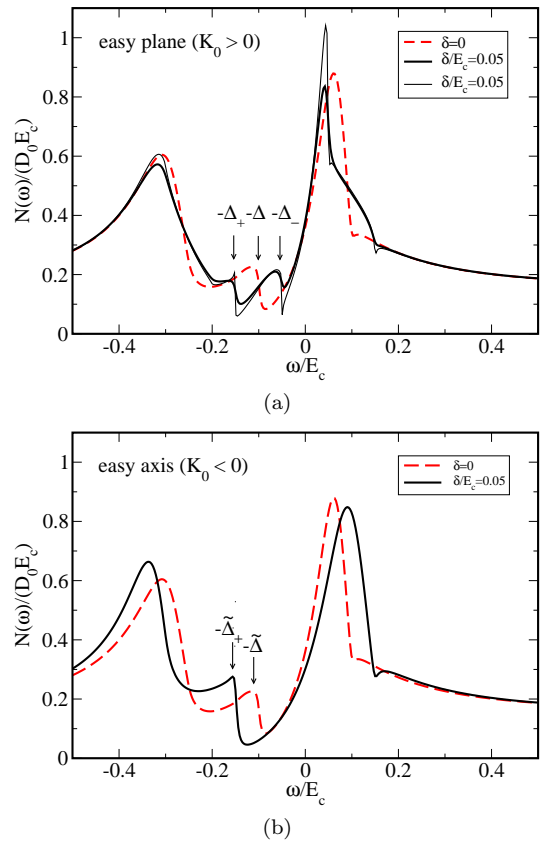

\centering
\subfigure[]
{\includegraphics[width=70mm,clip]{Fig9a}}
\subfigure[]
{\includegraphics[width=70mm,clip]{Fig9b}}
\caption{(Color online) Magnetic field effect on LDOS for easy-plane and easy-axis anisotropy $K_0$ of  impurity spin {\bf S} (a) $K_0>0:$  Two transitions to excited impurity states $|\pm 1\ra$ with split energies $\Delta_\pm=\Delta\pm\delta$ lead to field induced splitting of LDOS inelastic peak ($T=0$, $\mu/E_c=0.15, \Delta/E_c=0.1$).Thick and thin full lines correspond to numerical and analytical results respectively. Zero field case is shown as dashed line for comparison. (b) $K_0<0:$ only one inelastic transition to $|+1\ra$ excited state is present with shifted energy  $\tilde{\Delta}_+=\tilde{\Delta}+\delta$ appearing in the LDOS spectrum. Dashed line as before.
}
\label{fig:Fig9}
\end{figure}

First we discuss the easy-plane case ($K_0 >0$) as before which leads to the level scheme of Fig.~\ref{fig:Fig1}. 
We assume a symmetric splitting of the $|\pm1\ra$ excited doublet  by the field into Zeeman levels at $\Delta_\pm=\Delta\pm\delta$ where $\Delta =K_0 > 0$ and $\delta=g\mu_BH$. This corresponds to field along z-axis. In this case $\la S_z\ra$ is $\sim H^2$ and therefore the induced $U_1$ will be neglected. We also neglect the effect of Landau level splitting on the Dirac electrons for $g\mu_BH\ll E_c$. Therefore only the Zeeman splitting  of the impurity spin levels enters into the field dependence of the LDOS. It is shown in the top part of Fig.~\ref{fig:Fig9}a (full line) in comparison to the zero field case (broken line). The main effect is a splitting of the zero-field anomaly at $-\Delta$ into two well separated features at $-\Delta_\pm$ whose energy difference increases linearly with field. In addition the main resonance peaks are also slightly modified. 

Now we turn to the complementary easy-axis case ($K_0 <0$) where the level scheme of  Fig.~\ref{fig:Fig1} will be inverted with a doublet $|\pm 2\rangle$ ground state and first excited doublet $|\pm 1\rangle$. Both will split linearly in a field along the z-axis. Transitions from the ground state to the topmost singlet $|0\rangle$ have no dipole matrix element and do not appear in the self energy. Furthermore between the doublets only {\it one} transition $| +2\rangle \leftrightarrow |+1\rangle$ is relevant at $T=0$. Its energy depends on field strength according to $\tilde{\Delta}_+(H)=\tilde{\Delta}+\delta$ where now $\tilde{\Delta}=3|K_0| >0$ and $\delta$ as before. As a consequence only a single inelastic transition appears in the self energy of the easy-axis case. The corresponding LDOS is shown in Fig.~\ref{fig:Fig9}b and indeed exhibits only a single additional peak around  $-\tilde{\Delta}_+(H)$ that shifts away from the Dirac point with increasing field. To ensure comparison with Fig.~\ref{fig:Fig9}a  we rescaled the easy-axis parameters $M'\rightarrow M$ and $ \tilde{\Delta}\rightarrow \Delta$ (or $J_0^2\rightarrow (2/3)J_0^2$ and $|K_0|\rightarrow (1/3)|K_0|$).
 
The discussion of these two cases shows that the  splitting and shifting effect in the LDOS in a magnetic field could provide a spectroscopic tool to identify the spin states of the adsorbed magnetic impurity even in a very dilute limit. In particular the expected different LDOS signature for the easy - axis and -plane configurations may provide a tool to discriminate between them.

Finally we estimate the possible influence of the warping terms on the inelastic LDOS features. The lowest order warping energy appearing in the massless Dirac cone dispersion is \cite{thalmeier:11} $\epsilon_{w\bk}\simeq\lambda k_x(k_x^2-3k_y^2)$ . At the chemical potential (Fermi wave vector $k_F$) one has $\epsilon_{w_{k_F}}/\Delta \simeq (\mu/\Delta)(\mu/E_c)^2$. For parameters used here we then have $\epsilon_{w_{k_F}}/\Delta\simeq 10^{-2}$. Therefore averaging over cone states with different warping energies at  different \bk -vectors in the LDOS will not wipe out the inelastic structures at $\Delta$ discussed above. 

One should note that in reality in Bi$_2$Se$_3$ and Bi$_2$Te$_3$  the warping of Dirac half-cones is not symmetric and furthermore due to self-doping the chemical potential is far from the Dirac point \cite{hsieh:09} where warping effects are large. The chemical potential may be shifted close to the Dirac point by either carrier doping by
excess Se, Te  or slight chemical substitution of Bi by Ca or Sn \cite{hsieh:09} or application of a gate voltage \cite{chen:10b} or a combination of both \cite{checkelsky:11}. These methods also would have to be applied in the present context.

%
%

\section{Summary and conclusion}
\label{sec:conclusion}

We have investigated the effect of impurity spin states split by an anisotropy on the LDOS spectrum of topological insulators. In particular we analyzed a model relevant for Fe adatoms where the $S=2$ high spin state leads to a nonmagnetic singlet ground state. In this situation the usual first order elastic exchange scattering is absent. We have demonstrated, however, that in second order of the exchange interaction inelastic transitions to excited impurity spin states with energy $\Delta$ become important and have to be added to the usual nonmagnetic impurity potential scattering. The former lead to a dynamic self energy contribution in the T-matrix. It has singularities associated with the energy scale of inelastic transitions that leave its signature in the LDOS firstly by  modifying the position and widths of the bound state features in the spectrum already for chemical potential $\mu=0$ and secondly by introducing new anomalies in the LDOS directly connected with the energy of inelastic impurity spin excitations.  The position of the latter is independent of the chemical potential, however their prominence increases with chemical potential moving away from the Dirac point.

 Due to thermal occupation  the second order self energy is rapidly reduced when temperature becomes comparable to the inelastic transition energy leading to a rapid broadening of the bound state features and anomalies associated with $\pm\Delta$.
We have also suggested  that the temperature dependence of the LDOS spectrum in connection with its asymmetry gives direct information about the relative weight of pure potential scattering and second order exchange scattering since only the latter depends on temperature.

At temperatures $T\ll \Delta$ the inelastic exchange self energy anomalies in the LDOS at $\Delta$ may be used as a spectroscopic means to determine the surface impurity spin splitting microscopically rather than concluding it from indirect macrosocpic measurements as described in Sec~\ref{sec:costates}. This could be facilitated by application of a magnetic field which leads to a progressive splitting or shifting of  LDOS anomalies with increasing $H$, depending on the anisotropy type. In this way the inelastic exchange processes combined with variation  of $\mu$  and $H$ should make it possible to use LDOS measurements as a spectroscopic tool for the spin states of adatoms.
In particular this would be helpful in a situation when several distinct excitation energies $\Delta, \Delta' ..$ are accessible by exchange scattering which could be realized for surface adsorbed magnetic 4f impurities with their crystalline electric field splittings. In this case it would also be possible to achieve higher excitations energies comparable with the full range of the substitution- and gate-tuned chemical potential.\\

{\it Note added}: After completion of this manuscript we became aware of a related work on the LDOS of topological insulators in Ref. \cite{she:12}.

\appendix

\section{Matrix elements for helical state representation}
\label{sec:app1}

In this appendix we give the explicit form of the scattering matrix elements 
$\tU^{\tau\tau'}(\bk\bk')$ and $\tM^{\tau\tau'}_{\alpha\beta}(\bk\bk')$ in mean field and dynamical exchange 
Hamiltonian of Eq.~(\ref{eq:himpurityg}) , respectively, using the helical basis states.
The former are given by

\bea
\tU_0^{\tau\tau'}(\bk\bk')&=&
\left(\matrix
{A_+& ie^{-i\phi_\bk }A_- \cr
 ie^{i\phi_\bk}A^*_- &  A^*_-}\right)\no\\
 \tU_1^{\tau\tau'}(\bk\bk')&=&
\left(\matrix
{A_-& ie^{-i\phi_\bk }A_+ \cr
-ie^{i\phi_\bk}A^*_+ &  -A^*_-}\right)
\label{eq:amat}
\eea

with $A_\pm=1\pm\exp i(\phi_\bk-\phi_{\bk'})$ and \bk'=\bk+\bq.
The matrix elements in the remaining dynamical exchange term ${\cal H}_J$ are obtained as

\bea
\tM^{\tau\tau'}_{\alpha\beta}(\bk\bk')&=&\sum_{\sigma\sigma'}M_{\alpha\beta}^{\sigma\sigma'}V_{\sigma\sigma'}^{\tau\tau'}(\bk\bk')\no\\
\tensor{V}(\bk\bk')&=&P(\bk')^*\otimes P(\bk)
\label{eq:tm}
\eea

In general $\tM$ is a $(2S+1)^2\times4$ matrix where the row index $\alpha\beta$ denotes local spin excitations and the column index $\tau\tau'$ the transition between the helical half-cones. The phase factors resulting from helical state representation complicate the form of the matrix elements. However, as far as the LDOS at the impurity site for Dirac cones without warping is concerned, their effect is averaged out as discussed in Sec.~\ref{sec:scattering}.
The explicit form of the $\tensor{V}$-tensor in Eq.~(\ref{eq:tm}) in terms of the helical states phase factors, abbreviating $\phi=\phi_\bk, \phi'=\phi_{\bk'}$ is given by

\bea
\tensor{V}_{\bk\bk'}&=&\left(\matrix
{1 & ie^{-i\phi } &  -ie^{i\phi' } & e^{i(\phi'-\phi)} \cr
ie^{i\phi} &  1 & e^{i(\phi'+\phi)} & - ie^{i\phi'} \cr
- ie^{-i\phi'} & e^{-i(\phi'+\phi)} & 1 & ie^{-i\phi } \cr
 e^{-i(\phi'-\phi)} & - ie^{-i\phi'} &  ie^{i\phi} & 1}
 \right)\no\\
 \label{eq:vtens}
\eea

where the row index is $(\sigma\sigma')$ (z-componets of spin) and $(\tau\tau')$ (helicities) is the column index.

\section{Evaluation of second order exchange self energy}
\label{sec:app2}

In this appendix we evaluate the second order self energy due to the exchange scattering $\cal{H}_J$ in Eq.~(\ref{eq:himpurityc}).
From the diagram in Fig. \ref{fig:Fig1} it is obtained as
\bea
&&\Sigma_2(\bk,\tau,\bk'\tau';i\omega_n)= (J_0/2)^2P_\zeta\sum_{\alpha\beta; \bk'' \tau''}
\hM_{\bk\bk'}^{\tau\tau'}(\alpha\beta;\bk''\tau'')\no\\
&&\times \sum_{\omega_0}G(\bk'',\tau'',\omega_n-\omega_0)
\sum_{\omega_n}D(\beta,\omega_m)D(\alpha,\omega_m+\omega_0)\no\\
\label{eq:sig2a}
\eea
Here the total matrix element is defined by (see Eq.~(\ref{eq:tm}))
\bea
\hM_{\bk\bk'}^{\tau\tau'}(\alpha\beta;\bk''\tau'')=\tM_{\alpha\beta}^{\tau\tau''}(\bk\bk'')\cdot
\tM_{\alpha\beta}^{*\tau'\tau''}(\bk'\bk'')
\label{eq:hmat}
\eea
and the Dirac electron and CEF level pseudo-fermion Green's functions are given by
$G(\bk,\tau,\omega_n)=[i\omega_n-(E_{\bk\tau}-\mu)]^{-1}$ and $D(\alpha,\omega_n)=[i\omega_n-\zeta-E_\beta]^{-1}$ respectively with 
$\omega_n=(2n+1)\pi T$ and $\omega_0=2n_0\pi T$ denoting fermionic and bosonic Matsubara frequencies. Furthermore $\mu$ is the chemical potential and $\zeta$ is a Lagrange parameter to project out unphysical localized spin states. The projection is performed according to Ref.~\onlinecite{fulde:72} with $P_\zeta[\Sigma_2(\zeta)]=\lim_{\zeta\rightarrow\infty}Z^{-1}e^{\zeta/T}[\Sigma_2(\zeta)]$ where $Z=\sum_\alpha\exp(-E_\alpha/T)$ is the partition function for the split spin states of Fig.~\ref{fig:Fig1}. 
We obtain the expression for a general level scheme as
\bea
\Sigma_2(\bk,\tau,\bk'\tau';i\omega_n)&&=(J_0/2)^2\sum_{\alpha\beta;\bk''\tau''}
\hM_{\bk\bk'}^{\tau\tau'}(\alpha\beta;\bk''\tau'')\no\\
&&\times\frac{p_{\beta\alpha}[1-f(\epsilon_{\bk''\tau''})+n_B(\Delta_{\alpha\beta}]}
{i\omega_n-(\epsilon_{\bk''\tau''}+\Delta_{\alpha\beta})}
\label{eq:sig2b}
\eea
Here $n_\alpha=Z^{-1}\alpha\exp(-E_\alpha/T)$ are the thermal occupation factors of local spin states and $p_{\beta\alpha}=n_\beta-n_\alpha$ their differences. The Fermi and Bose functions are denoted by $f(\omega)$ and $n_B(\omega)$, respectively. Furthermore $\Delta_{\alpha\beta}=E_\alpha-E_\beta$ are their excitation energies and $\epsilon_{\bk''\tau''}=E_{\bk''\tau''}-\mu$ quasiparticle energies with respect to the chemical potential. 


\bibliography{paper}

\end{document}